\documentclass[preprint,  amssymb, amsmath, superscriptaddress] {revtex4-1}
\usepackage{hyperref}
\usepackage{graphicx}
\begin{document}

\title{Origin of the giant linear magnetoelectric effect in perovskite-like multiferroic BiFeO$_3$}

\author{A.~F.~Popkov}
\affiliation{Moscow Institute of Physics and Technology~(State University), 141700, Dolgoprudny, Russia}
\affiliation{National Research University of Electronic Technology (MIET), pas. 4806, bld. 5, Zelenograd, Moscow, Russia}

\author{M.~D.~Davydova}
\email{davydova@phystech.edu}
\affiliation{Moscow Institute of Physics and Technology~(State University), 141700, Dolgoprudny, Russia}

\author{K.~A.~Zvezdin}
\email{konstantin.zvezdin@gmail.com}
\affiliation{Moscow Institute of Physics and Technology~(State University), 141700, Dolgoprudny, Russia}
\affiliation{Prokhorov General Physics Institute, Russian Academy of Sciences, 119991, Moscow, Russia}

\author{S.~V.~Solov'yov}
\affiliation{National Research University of Electronic Technology (MIET), pas. 4806, bld. 5, Zelenograd, Moscow, Russia}

\author{A.~K.~Zvezdin}
\affiliation{Moscow Institute of Physics and Technology~(State University), 141700, Dolgoprudny, Russia}
\affiliation{Prokhorov General Physics Institute, Russian Academy of Sciences, 119991, Moscow, Russia}

\date{\today}
\begin{abstract}
In this article the mechanism of the linear magnetoelectric (ME) effect in the rhombohedral multiferroic BiFeO$_3$ is considered. The study is based on the symmetry approach of the Ginzburg-Landau type, in which polarization, antiferrodistortion, and antiferromagnetic momentum vectors are viewed as ordering parameters.  We demonstrate that the linear ME effect in BFO is caused by reorientation of the antiferrodistortion vector in either electric or magnetic field. The numerical estimations, which show quantitative agreement with the results of the recent measurements in film samples,  have been performed. A possibility of significant enhancement of the magnetoelectric effect by applying an external static electric field has been investigated. The considered approach is promising for explaining the high values of the ME effect in composite films and heterostructures with BFO.
\end{abstract} 	


\maketitle

	Magnetoelectric (ME) effect opens up wide prospects for practical applications of multiferroic materials in various fields of nanoelectronics, microwave- and optoelectronics \cite{Birola, Rivera2009, Fiebig, Spaldin, Schmid, Zvezdin2012}. Intense investigation of the magnetoelectric effect dates back to the 2000-s, with the advent of perspective film oxide perovskite multiferroics ABO$_3$. The pseudocubic multiferroic BiFeO$_3$  (BFO) stands out among other materials, since it has both ferroelectric and magnetic ordering at room temperature \cite{Eerenstein, Catalan}. However, for a long time the linear magnetoelectric effect in BFO was not found. The bismuth ferrite is an antiferromagnet with the structure, which is not fully G-type, but shows a cycloidal spin structure with a period of 62 nm\cite{Sosnowska1982, Sosnowska1995}. Canting of the Fe$^{3+}$ sublattices leads to a weak local magnetization, which averages to zero over a period of the spin cycloid structure \cite{Kadomtseva2004}. Due to the spin cycloid structure, the volume average of the linear magnetoelectric effect also equals to zero. Therefore, firstly only the quadratic ME effect had been observed, and the value of the obtained magnetoelectric susceptibility tensor elements  of the order of 10$^{-19}$ s\,A$^{-1}$ had been found \cite{Tabares-Mun̄oz}. In 2003, the article \cite{Wang2003} by Wang \textit{et al} was published, in which a giant value of magnetoelectric coefficient of an order of 3 V/(cm\,Oe) was found in heteroepitaxially constrained thin BFO films, and the interest in the bismuth ferrite has been revived. The spin cycloid structure can be destroyed and the linear ME effect can be recovered by applying an external magnetic field of a large magnitude \cite{Popov1993, Popov1994, Popov2001, Ruette2004} and using chemical doping \cite{Murashov,Bras, Gabbasova}. In recent years, more experimental studies have been conducted, which indicate the giant magnetoelectric effect, enhanced in composites and multilayer heterostructures based on BFO \cite{Bai, Prosandeev, Lorenz2014, Kumar}.  Latest studies report the value of the linear magnetoelectric coefficient of approximately 1.6 V/(cm\,Oe) in bulk BFO \cite{Caicedo}, 4.2 V/(cm\,Oe) in BiFeO$_3$ films and 24 V/(cm\,Oe) in composite films with BFO \cite{Lorenz}. However, the origin of the large values of the ME effect observed by numerous experiments in BFO remains unexplained, and the attempts to provide theoretical grounding to it have failed.   
	 
Pioneering works introduced atomistic-like approach, where firstly the value of intristic ME coefficient of the order of 10$^{-2}$ V/(cm\,Oe) was obtained \cite{KornevLOW}, which was close to experimental values \cite{Rivera97} known at that time, but is much lower than the values observed in the recent experiments. As well, in \cite{Lisenkov}, using a similar method, it has been concluded that the magnetoelectric properties of BFO can be explained without resorting the linear ME coefficients. In the light of the latest experiments it has become evident that these results need reconsideration. To the best of our knowledge, there is no theoretical publications that consider the linear ME in BFO and give the value of the ME coefficient close to experimental.

\begin{figure}[h]
    \includegraphics[width =0.45\columnwidth]{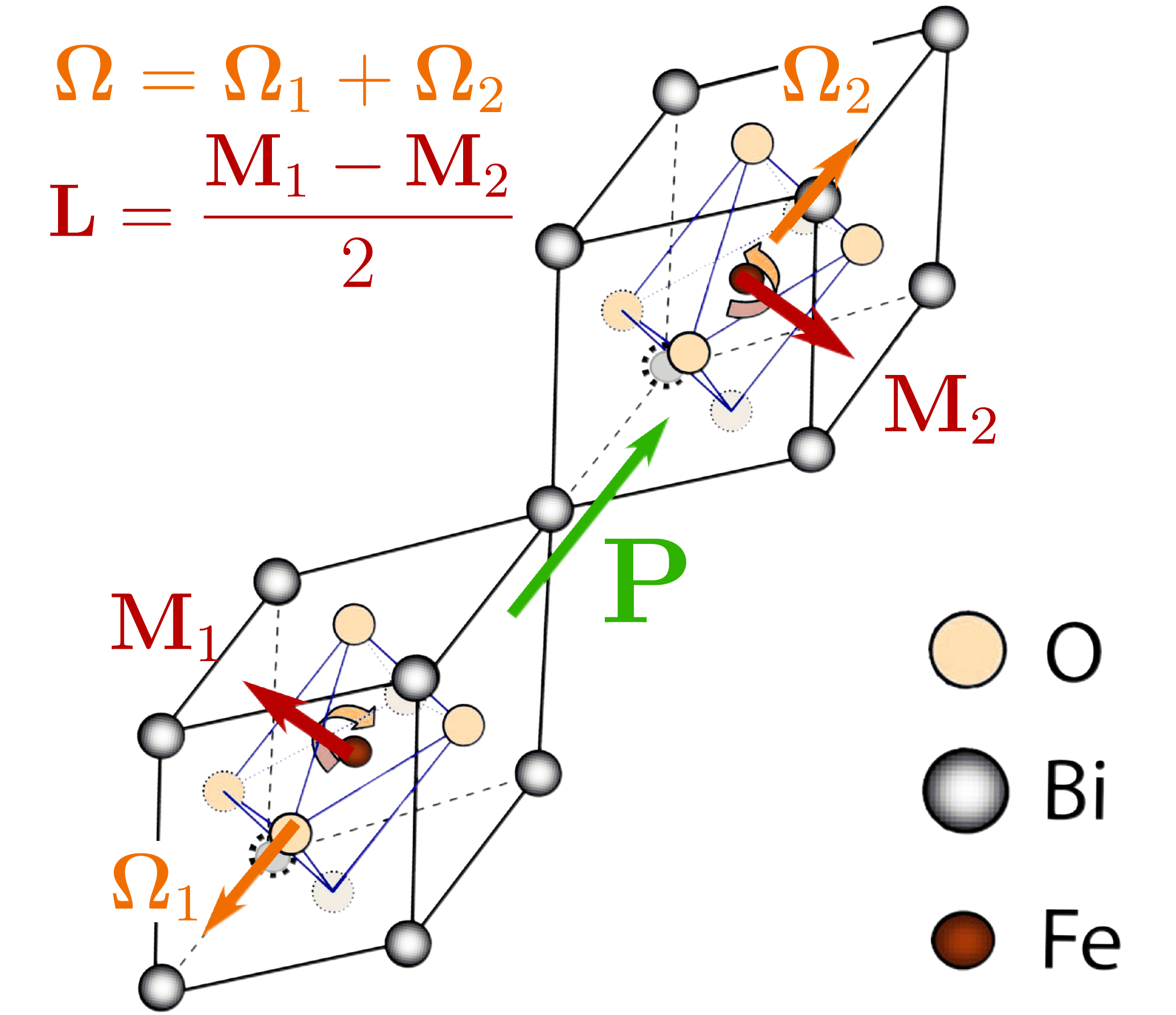}
    \caption{ Bismuth ferrite rombohedrally distorted perovskite cell doubled by antiparallel rotation of oxygen octahedra.}
    \label{fig1}
\end{figure}

 In our study we provide theoretical background and explain the large value of the intrinsic linear ME coefficient of BFO.  The obtained value of the linear ME coefficient is close to the maximum measured value in recently published articles \cite{Lorenz}. We demonstrate that the antiphase oxygen octahedra rotation is responsible for occurrence of  magnetoelectricity in BFO.  Our approach is based on the symmetry representation of the thermodynamic potential in the Ginzburg-Landau approach. We use an invariant expansion of the thermodynamic potential in powers of the ordering parameters, namely antiferrodistortion ($\boldsymbol{\Omega}$), polarization ($\mathbf{P}$), and antiferromagnetic ($\mathbf{L}$) vectors. In Fig. \ref{fig1} the rhombohedral perovskite-type doubled unit cell is shown with corresponding illustrations of order parameters. The doubling of the unit cell of the crystal structure occurs due to the antiphase rotation of the oxygen octahedra, which surround the Fe$^{3+}$ ions (antiferrodistortion). Displacement of the oxygen and iron ions within the double cell is responsible for the spontaneous polarization (see Fig. \ref{fig1}). 
 
There are two alternative ways to derive the tensor of the linear magnetoelectric effect $\hat \alpha$  (the magnetoelectric tensor): either using the definition $\alpha_{ij} = 4 \pi \frac {\delta  P _i}{\delta H_j }$ or the definition $\alpha_{ji} = 4 \pi \frac {\delta M _i}{\delta E_j }$, where $\mathbf M =(\mathbf M_1 + \mathbf M_2)/2$ is the  magnetic moment per formula unit,  $\mathbf M_1$ and $\mathbf M_2$ are the magnetic moments of the two iron atoms in the aligned unit cell (see Fig. \ref{fig1}).
In our work, we find the tensor $\hat \alpha$ through modulation of the  magnetic moment $\mathbf M$ in the external electric field $\mathbf E$. Due to weak ferromagnetizm of bismuth ferrite, the spontaneous magnetic moment can be expressed as  $\mathbf M = \chi_\perp \mathbf H_{D}$, where $\chi_\perp$ is the component of the magnetic susceptibility, which is perpendicular to the antiferromagnetic vector $\mathbf L = (\mathbf M_1 - \mathbf M_2)/2$  (see Fig. \ref{fig1}), and $\mathbf H_{D}$ is the Dzyaloshinskii field. The latter is defined by the expression $\mathbf H_{D} = \mathbf D \times \mathbf L$, where $\mathbf D$ is the Dzyaloshinskii vector \cite{Dzyaloshinskii}. The direction of the Dzyaloshinskii vector coincides with the direction of the antiferrodistortion vector $\boldsymbol \Omega$ \cite{Ederer2005}. Later, it has been shown \cite{ZvezdinD} that this dependency can be written as $\mathbf D = K \boldsymbol \Omega$, where $K=M_S/( \chi_\perp  \Omega_0 M_0)$, $M_S$ denotes the equilibrium magnetization, $M_0$ is the magnetic moment of one irom atom in the unit cell, and $\Omega_0$ is the magnitude of the AFD vector  in the equilibrium state. All the physical values mentioned can be determined from experiment. Therefore, using the equations above and  introducing the normalized  antiferromagnetic   vector $\mathbf l = (\mathbf M_1 - \mathbf M_2)/ 2 M_0 $, we express the magnetoelectric tensor with the following formula:
\begin{equation} \label{alpha_def}
 	\alpha_{ji} =  \frac{4 \pi  M_S}{\Omega_0}\epsilon_{inm} \frac {\delta \boldsymbol \Omega_n}{\delta E_j } l_m,
 	\end{equation}
Thus we have shown that the mechanism of the occurrence of the ME effect can be interpreted using the change of the AFD vector $\boldsymbol \Omega$ in the external electric field. Namely, one has to find the value of the tensor of the electric susceptibility of the antiferrodistortion $\eta_{ij} = \delta \Omega_i/ \delta E_j $ in order to obtain the magnetoelectric tensor from \eqref{alpha_def}.

  For this purpose we introduce the ferroelectric part of the thermodynamic potential $\Phi_{st}(\boldsymbol{\Omega},\mathbf{P},\mathbf{E}) $. 
   In the coordinate system  $Ox  \parallel [ 001]$,  $Oy  \parallel [  010]$, $Oz  \parallel [100]$  the ferroelectric part of the thermodynamic potential has  following form \cite{PRB_new}:
\begin{align}\label{pot_st}
\begin{split}
&\Phi_{st}(\boldsymbol{\Omega},\mathbf{P},\mathbf{E}) = 
 a_1(P_x^2 + P_y^2 + P_z^2)  + a_{11}(P_x^4 + P_y^4 + P_z^4) \\
&+a_{12}(P_x^2P_y^2 + P_y^2P_z^2 + P_z^2P_x^2)  - \mathbf{P}\mathbf{E}
+ b_1(\Omega_x^2 + \Omega_y^2 + \Omega_z^2)
\\
&+ b_{11}(\Omega_x^4 + \Omega_y^4 + \Omega_z^4)
 +b_{12}(\Omega_x^2\Omega_y^2 + \Omega_y^2\Omega_z^2 + \Omega_z^2\Omega_x^2) 
 \\
&- t_{11}(\Omega_x^2P_x^2 + \Omega_y^2P_y^2 + \Omega_z^2P_z^2)
\\
&-t_{12}\left ( \Omega_x^2(P_y^2+P_z^2)+  \Omega_y^2(P_x^2+P_z^2) + \Omega_z^2(P_x^2+P_y^2)\right )
\\
&- t_{44}(\Omega_x P_y \Omega_y P_x + \Omega_y P_z \Omega_z P_y + \Omega_z P_x \Omega_x P_z),
\end{split}
\end{align}
\noindent where $a_{1}$, $b_{1}$, $a_{ij}$, $b_{ij}$, and $t_{ij}$  are parameters that are defined later in the Letter.

 The dependences of the equilibrium order parameters $\boldsymbol{\Omega}_0$  and  $\mathbf{P}_0$ on the applied static external magnetic and electric fields $(\mathbf H_0,\mathbf E_0)$ are found by minimization of the thermodynamic potential of the crystal $\Phi$ using the variational equations $\delta \Phi(\mathbf H_0,\mathbf E_0) / \delta \mathbf{P} = 0$  and  $\delta \Phi(\mathbf H_0,\mathbf E_0) / \delta \boldsymbol{\Omega} = 0$.
 After that, we consider the small deviation of the external electric field: $\mathbf E = \mathbf E_0 + \delta \mathbf E$. In the linear approximation the polarization can be naturally represented as $ \mathbf{P} =\mathbf{P}_0  + \hat \kappa \delta \mathbf{E} $, where  $\hat \kappa$ is the electric susceptibility tensor, and $\mathbf{P}_0=\mathbf{P}(\mathbf H_0,\mathbf E_0)$ is the equilibrium  polarization. We introduce the same representation for the AFD vector: $  \boldsymbol{\Omega} =\boldsymbol{\Omega}_0 +\hat  \eta \delta \mathbf{E}$, $\boldsymbol{\Omega}_0=\boldsymbol{\Omega}(\mathbf H_0,\mathbf E_0)$ is the equilibrium antiferrodistortion vector. Below we consider the bismuth ferrite in absence of external fields (unperturbated crystal), $(\mathbf E_0, \mathbf H_0) =(0,0)$. In this case the equilibrium ordering parameters  $\mathbf P_0$, $\boldsymbol{\Omega}_0$ $\parallel (1,1,1)$. Now we consider the variation of the thermodynamic potential due to the deviation of the external electric field. 
 We introduce three tensors $A_{ij} = \frac{\delta^2\Phi }{ \delta P_i  \delta P_j }$, $B_{ij} = \frac{\delta^2\Phi }{ \delta \Omega_i  \delta P_j }$ and $C_{ij} = \frac{\delta^2\Phi }{ \delta \Omega_i  \delta \Omega_j }$, which describe the quadratic form of the expansion of the thermodynamic potential at the  point of equilibrium. 
The variational equations may be rewritten as following: 
\begin{align}\label{deviations}
\begin{split}
 &A_{ij} \delta P_j + B_{ij} \delta \Omega_j  =-\frac{\delta^2\Phi }{   \delta P_i \delta E_j} \delta E_j, \\
&B_{ij} \delta P_j +C_{ij} \delta \Omega_j  =-\frac{\delta^2\Phi }{   \delta \Omega_i \delta E_j} \delta E_j.
\end{split}
\end{align}

The three introduced tensors have similar structure, namely $A_{ij} = A_1  \delta_{ij}+ A_2 ( 1- \delta_{ij})$ ($ \delta_{ij}$ is the Kronecker symbol), where 
\begin{align}\label{TPtensors}
\begin{split}
 &A_1 =2a_1+\left (4a _{11}+\frac{4}{3} a_{12} \right )P_0^2 - \left(\frac{2}{3} t_{11} +\frac{4}{3} t_{12} \right)\Omega_0^2, 
 \\
 &A_2= \frac{4}{3}a_{12}P_0^2-\frac{1}{3}t_{44}\Omega_0^4 
 \\
  &B_1 =-\frac{4}{3}t_{11}P_0 \Omega_0 -\frac{2}{3}t_{44}P_0 \Omega_0, 
  \\ 
  &B_2 = -\frac{4}{3}t_{12}P_0 \Omega_0 -\frac{1}{3}t_{44}P_0 \Omega_0.
  \\
  &C_1 =2b_1+\left (4b _{11}+\frac{4}{3} b_{12} \right )\Omega_0^2 - \left(\frac{2}{3} t_{11} +\frac{4}{3} t_{12} \right)P_0^2, 
  \\
   &C_2 = \frac{4}{3}b_{12}\Omega_0^2-\frac{1}{3}t_{44}P_0^2.
\end{split}
\end{align}

Given that $\delta^2\Phi /   \delta P_i \delta E_j  = - \delta_{ij}$ and $\delta^2\Phi /\delta \Omega_i \delta E_j  =  0$,  we obtain the following expressions for tensors of the linear expansion:
\begin{align}
 &\kappa_{ik} =  \Delta ^{-1}_{ij}  C_{jk}, \ &\eta_{ik} &= \Delta ^{-1}_{ij}   B_{jk} \label{dev_tensors2},
\end{align}
where  $\hat \Delta=\hat A\hat C - \hat B^2$.

Note, that the magnetoelectric tensor \eqref{alpha_def} can be expressed using tensor $\hat \eta$, for which is true $\eta_{ij} = \delta  \Omega_i / \delta  E_j$. We find the last tensor using \eqref{dev_tensors2}, and transfer to the ''rhombohedral'' coordinate system with $Ox  \parallel [ 1 1 \bar 2]$,  $Oy  \parallel [  1 \bar 1 0]$, $Oz  \parallel [ 1 1 1]$. In this system the tensor $\hat \eta$ has diagonal form:
\begin{equation} \label{eta_def}
\hat \eta =\begin{pmatrix}
\eta_{\perp} & 0 & 0\\ 
0 & \eta_{\perp} & 0\\ 
0 & 0 & \eta_{\parallel}
\end{pmatrix},
\end{equation}
where $\eta_{\perp}$  and  $\eta_{\parallel}$ characterize the electric susceptibility of the AFD vector in the longitudinal and the transverse directions in relation to the unit vector $\mathbf e_P = \mathbf P_0 /P_0$.  These components are defined by the following equations:
\begin{align}\label{eta_comp}
\begin{split}
\eta _\parallel &= \frac{((\mu_1 - \nu _1)-(\mu_2 - \nu_2))(B_1+B_2)}{(\mu_1 - \nu _1)(\mu_1 - \nu _1+\mu_2-\nu_2)-2(\mu_2-\nu_2)^2},
\\
\eta _\perp &=\frac{((\mu_1 - \nu _1)+2(\mu_2 - \nu_2))(B_1-B_2)}{(\mu_1 - \nu _1)(\mu_1 - \nu _1+\mu_2-\nu_2)-2(\mu_2-\nu_2)^2},
\end{split}
\end{align}
where $\mu_1 = A_1 C_1 + 2 A_2 C_2$, $\mu_2 = A_1 C_2 + A_2 C_1 + A_2 C_2$, $\nu_1 = B_1^2 + 2 B_2^2$,  $\nu_2 = 2 B_1 B_2 + B_2^2$. For reference we present the expression for the tensor $\hat \eta$ in the pseudocubic coordinate system:
\begin{equation} \label{eta_qubic}
 \eta_{ij} =  \eta_1  \delta_{ij} + \eta_2 ( 1 -  \delta_{ij}),
\end{equation}
where  $\eta_1 = \frac{1}{3}(\eta_{\parallel}+2\eta_{\perp})$ and $\eta_2 =\frac{1}{3}( \eta_{\parallel}-\eta_{\perp})$.

	Thus, we obtained the magnetoelectric tensor in the coordinate system $O\tilde{x} \parallel [1\bar10]$, $O\tilde{y} \parallel [11\bar2]$, and  $O\tilde{z} \parallel [111]$   using  the definition of  $\hat \eta$, \eqref{alpha_def} and \eqref{eta_def}:
	\begin{equation} \label{result_alpha}
\hat \alpha= 4 \pi \chi_\perp \frac{H_{D}^0}{\Omega_0}\begin{pmatrix}
0 & \eta_\perp l_z & -\eta_\perp l_y\\ 
-\eta_\perp l_z & 0 & \eta_\perp l_x\\ 
 \eta_\parallel l_y & \eta_{\parallel} l_x& 0
\end{pmatrix}
\end{equation}
This tensor corresponds well to the result from \cite{Popov1993}, where derivation of the ME tensor was based on the symmetry properties of the rhombohedral ferroelectric phase of the BFO. We should note, that proceeding from the alternative definition of the ME tensor $\alpha_{ij} = 4 \pi \frac {\delta  P _i}{\delta H_j }$, one may use the same approach to find the change of the polarization in the external magnetic field, and obtain the same result for the ME tensor.

Our numerical calculations are based on the thermodynamic potential \eqref{pot_st} with parameters $a_{i}$, $b_{i}$,$a_{ij}$, $b_{ij}$, and $t_{ij}$, which are present in Table \ref{table1}. The values of these parameters have been selected  \cite{Kulagin}
 to fit to the available experimental data (see, for example, \cite{Kadomtseva2004}) and \textit{ab initio} calculations \cite{Lisenkov2009}.  
  The values of the ME tensor can be estimated as $\alpha_\parallel = 4 \pi \frac {M_{S}} {\Omega_0}|\eta_\parallel|$ and  $\alpha_\perp =  4 \pi \frac {M_{S}}{\Omega_0}|\eta_\perp|$. We assume that  $M_S = H_D^0 \chi_\perp$, $H_D^0 \approx 1.4 \times 10^5$ Oe is the magnitude of the  Dzyaloshinskii field in the equilibrium, $\chi_\perp \approx 5 \times 10^{-5}$, and $\Omega_0=0.21$,  and obtain $\alpha_\perp\approx 2.2\times 10^{-3}$   (in gaussian units), or  $0.67$ V/(cm\,Oe), and $\alpha_\parallel \approx 3.27\times  10^{-2}$   (in gaussian units), or  $9.81$ V/(cm\,Oe) (with accuracy up to the components of the normalized vector $\mathbf l$, $|\mathbf l| \approx 1$). The order of the magnitude of the last value is consistent with the value of the linear magnetoelectric effect $\alpha \approx 1.4\times 10^{-2}$   (in gaussian units), present in \cite{Lorenz}. It is more than an order of magnitude greater than the value of the ME coefficient  $\alpha_\parallel \approx 1.24\times  10^{-3}$  in the well-known multiferroic Cr$_2$O$_3$ \cite{Wiegelmann, Folen}, but is less than the effect in TbPO$_4$ giving 720 ps/m  (T = 1.50 K) \cite{Rivera2009}. We should note that the obtained result is suitable for a single-domain sample, while for a multi-domain sample the the ME effect would be smaller.
  
\noindent
\squeezetable
\begin{table}
\caption{\label{table1} Numerical values of the parameters in thermodynamic potential \eqref{pot_st} of BFO, which have been used in calculations. }
\begin{ruledtabular}
\begin{tabular}{ccccccccc}
  $a_1$(J\,m/C$^2$) & $a_{11}$(J\,m$^5$/C) & $a_{12}$(J\,m$^5$/C$^4$) & $b_1$(J/m$^3$) & $b_{11}$(J/m$^3$) & $b_{12}$(J/m$^3$)  & $t_{11}$(J\,m/C$^2$)  & $t_{12}$(J\,m/C$^2$)  &$t_{44}$(J\,m/C$^2$)  \\
\hline
$-8.05\times10^7$ & $5.22\times10^7$ & $6.87\times10^7$ & $-3\times10^8$ & $1.3\times10^9$ & $1.9\times10^9$&$-2.6\times10^8$ & $-2.5\times10^8$  & $5\times10^7$\\
\end{tabular}
\end{ruledtabular}
\end{table}
As an illustration of applications of our approach for investigation of mechanisms of enhancement of the linear ME effect, let's consider the following idea.
Due to \eqref{alpha_def} and \eqref{result_alpha} the giant values of the ME effect are expected to occur when the derivatives of the AFD vector with respect to the electric field are experiencing critical behavior. In particular, the studies of the electric field-induced structure and magnetic changes in BFO \cite{Kulagin, Lisenkov} have shown that at certain magnitudes of the external electric field a reorientation of the antiferromagnetic structure occurs. In the vicinity of the critical fields the derivatives of the AFD vector components may approach to infinity, which, in turn, leads to an unlimited increase of the magnetoelectric effect in theory. The BFO films  in experiments are often oriented perpendicularily to [001] \cite{Heron,Palai, Sando}, therefore as an example we consider the case of a phase trantision in $\mathbf E \parallel [001]$.

\begin{figure}[b]
    \includegraphics[width =\columnwidth]{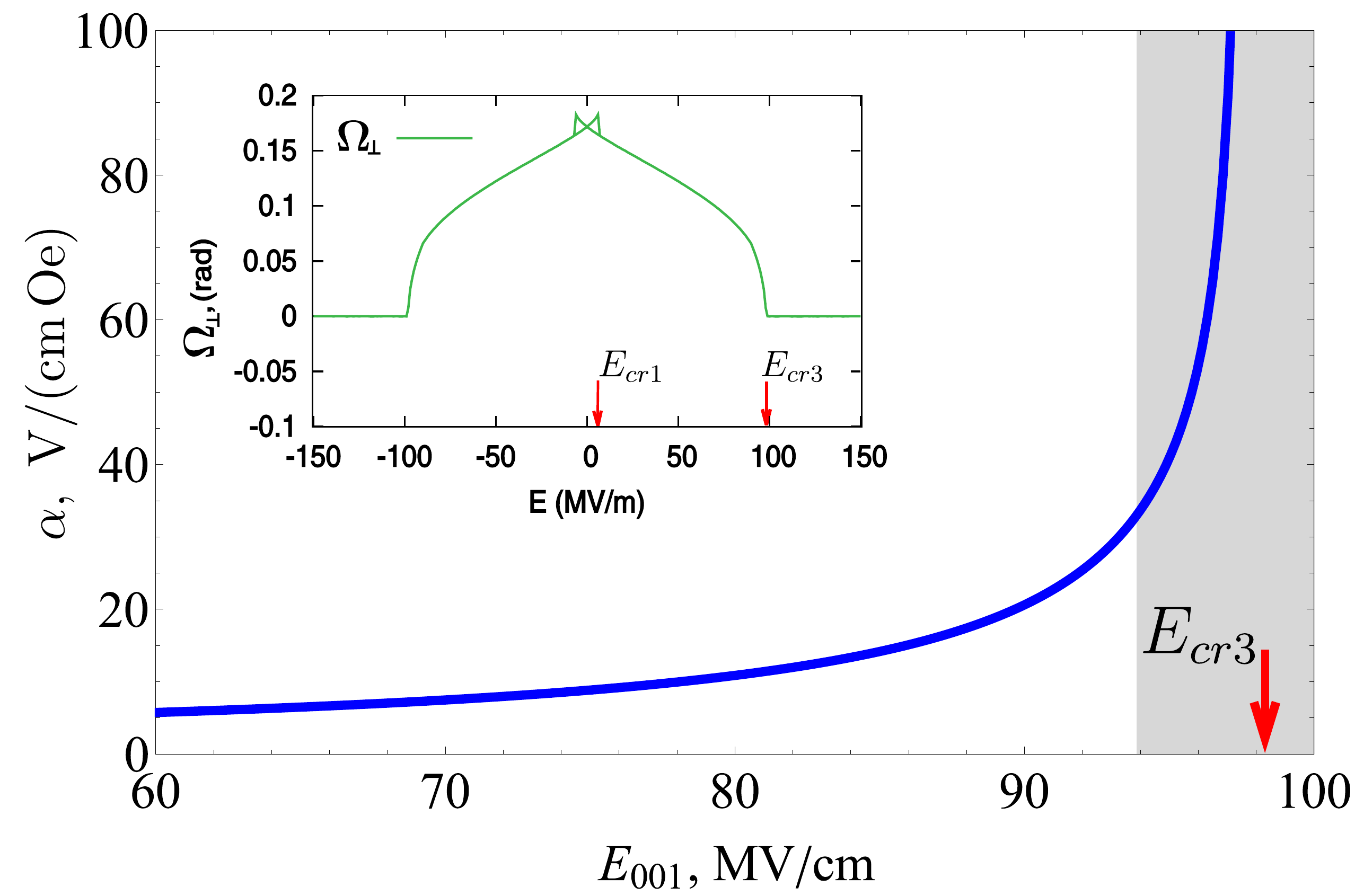}
    \caption{Dependence of the ME coefficient $\alpha$ on the external electric field, which is applied in $[001]$ direction. Insertion:  dependence of the $\Omega_\perp$  on the applied external electric field. }
    \label{fig2}
\end{figure}

	 In the external electric field  $\mathbf E \parallel [001]$ there are three critical points, namely $E_{cr1}$,  $E_{cr2}$ and  $E_{cr3}$, at which phase transitions occur \cite{Kulagin, PRB_new}.  Here we present an example of analysis of the asymptotic behavior of the ME tensor near $E_{cr3}$ (see the insertion in Fig. \ref{fig2}).  At the large values of the electric field ($E>E_{cr3}$) polarization is aligned with the direction of the electric field ($ P = P_\parallel$), and the AFD vector consists only of the perpendicular component $\Omega_\perp$. We assume that $\Omega_\perp$  is parallel to the $ [110]$ direction, which is not the only direction due to symmetry. In this case the thermodynamic potential in the coordinate system  $Ox\parallel [001]$, $Oy\parallel [010]$, $Oz\parallel [100]$ acquires the following form:
	\begin{equation} \label{pot_simpl}
\Phi(\boldsymbol{\Omega},\mathbf{P},\mathbf{E})=\alpha_1P_\parallel^2+\alpha_{11}P_\parallel^4-P_\parallel E_{001}+\beta_1\Omega_\perp^2+(2\beta_{11}+\beta_{12})\Omega_\perp^4-t_{12}P_\parallel^2 \Omega_\perp^2
\end{equation}

At $E_{cr3}$ a phase transition of the second order occurs: the expression standing in front of the term $\Omega_\perp^2$  changes its sign, and the coefficient in front of $\Omega_\perp^4$ remains positive. Solving the minimization problem in the approximation of the small value of  $\Omega_\perp$, we obtain the following expressions near the equilibrium point:
$P_\parallel \approx P_\parallel^0 + \gamma_P \Delta E_{001}$, $ \Omega_\perp \approx \gamma_\Omega \sqrt \frac {\Delta E_{001}} {E_{cr3}}$,
where  $P_\parallel^0\approx 1.1$ C/m$^2$,   $\gamma_P \approx -1.69 \times 10^{-9}$ C$^2$/(J$\,$m), $\gamma _\Omega \approx 0.21$ rad, the deviation of the electric field magnitude from the critical value is $\Delta E_{001} = E_{cr3} - E_{001}$, $E_{cr3}
\approx 9.81\times 10^7$ V/m. 
For estimation of the ME tensor components using \eqref{alpha_def}, we introduce the magnetoelectric coefficient $\alpha $, which is proportional to the elements of the tensor $\hat \alpha$ with accuracy up to the components of the vector $\mathbf l$. The following expression indicates the unlimited growth of the ME effect:
           \begin{equation} \label{critical_2}
\alpha \propto \frac{\partial \Omega_\perp}{\partial E_{001}}= -\frac{\gamma_\Omega}{2\sqrt{E_{cr3}}}\cdot\frac{1}{\sqrt{E_{cr3}-E_{001}}}.
\end{equation}

The same asymptotic behavior may be obtained near other critical points. In Fig. \ref{fig2} is shown the dependence of the magnetoelectric coefficient $\alpha$ on the electric field $E_{001}$. 
Thus we have shown that the search of the similar setups, when the derivatives of the AFD vector with respect to the electric field components become considerably large, may lead to significantly enhanced results for the magnetoelectric effect in BFO.

In order to ensure that the system is thermodynamically stable \cite{Brown} the next condition must be satisfied:
  	           \begin{equation} \label{brown}
\alpha _{ij} < 4 \pi \sqrt{ \chi_{ii} \kappa_{jj}},
\end{equation}
where  $\hat \chi $ and  $\hat \kappa$ are magnetic and  electric susceptibility tensors correspondingly.  For the upper boundary of the value of the $\alpha _{ij}$, at which the condition \eqref{brown} is still satisfied, approximate estimation gives the order of 10$^{-1}$ gaussian units. This sets the limit to the external electric fields in our approach (the limitation area is denoted by gray filling in Fig. \ref{fig2}). 

In conclusion, we have shown that the nature of the magnetoelectric effect in BiFeO$_3$ lies in the orientation change of the AFD vector in the external electric field.  We note that the developed model for the magnetoelectric interaction in BFO, indicates the prospect of the ME effect enhancement. The enhancement mechanism may lie in the softening of the mode of the reorientation of the antiferrodistortion vector, for example, under the elastic stresses in film heterostructures. Also, our investigations have shown that in presence of the external electric field of critical magnitude (at which еру phase transition of the second order occur), the giant ME effect may be observed. Our study may be relevant, in particular, for explanation of recent experimental observations of the growth of the ME effect in thin-film heterostructures with BFO \cite{Wang2003, Liu2013, Prosandeev, Lorenz2014, Lorenz}, and for search for tools of enhancement of the linear magnetoelectric effect in BFO.

This work was supported by the Russian Foundation for
Basic Research (No. 13-07-12405 ofi-m2, No. 14-02-91374 ST-a, No. 13-07-12443 ofi-m2)
and 50 Labs Initiative of Moscow Institute of Physics and
Technology (MIPT).

%

\end{document}